\documentclass[a4paper]{article}

\usepackage{icrc2013}
\usepackage{amsmath}
\usepackage{float}

\title{Study Of Casleo Clear Sky Aerosol Loads In 2011 From One Year Of Aeronet Quality Assured Data}

\shorttitle{Study Of Casleo Clear Sky Aerosol}

\authors{
Lidia Otero$^{1}$,
Pablo Ristori$^{1}$,
D'El\'{i}a Ra\'{u}l$^{1}$,
Juan Pallotta$^{1}$,
Eduardo Quel$^{1}$
for the CTA Consortium.
}

\afiliations{
$^1$ UNIDEF-CEILAP-(CITEDEF-CONICET), UMI-IFAECI-CNRS (3351) - Buenos Aires, Argentina. \\
}

\email{lotero@citedef.gob.ar}

\abstract{In this work we analyze one year observation of an Aeronet (GSFC-NASA Aerosol Robotic Network) sun-photometer installed on January 11, 2011 in CASLEO and being operational up to date. The main goal of placing the instrument in this location is to characterize the aerosol loads of this astronomical complex which is close and has the same sky characteristics of El Leoncito (31$^\circ$ 43,33' South - 69$^\circ$ 15,93' West, 2552 m ASL) one of the southern candidate site for Cherenkov Telescope Array (CTA). The low aerosol optical depth (AOD) annual mean of 0.038 measured at 500 nm shows exceptional clear sky quality. Data is compared with the measurements being done at Mauna Loa (19$^\circ$ 32.34' North, 55$^\circ$ 34.68' West, 3397 m ASL), where Aeronet reference instruments are being re-calibrated two to four times per year. Long term MODIS observations are studied, showing that the site is far enough to biomass burning transport regions to be affected by its influence.}

\keywords{lidar, aerosols, atmosphere, Aeronet.}

\begin{document}
\maketitle

%Begin a section.
\section{Introduction}

\subsection{The Observatory}

The astronomical complex “El Leoncito” or Complejo Astronómico “El Leoncito” (CASLEO) started its operations on September 12, 1986. It is placed in the Province of San Juan at the Calingasta Department at 40 km from El Barreal inside the El Leoncito National Park. It has many facilities as the optical telescopes “Jorge Sahade”, “Helen Sawyer Hogg”, “Horacio Ghliemetti”, an astrograph, and a sub-millimeter solar telescope. Further information about this instrumentation can be found in \cite{bib:casleo}. Due to its excellent sky quality and infrastructure this place was chosen as one of the two Argentinean candidate sites for the CTA Southern Observatory (site location: 31$^\circ$41’19”S and 69$^\circ$16’18”W).

\subsection{The instruments}

Aeronet is a ground base remote sensing aerosol monitoring network created by NASA to support NASA, CNES and NASDA Earth’s satellite systems based on weather resistant sun and sky scanning radiometers and a standardized calibration and data processing protocol with freely available observation \cite{bib:aeronet}. Its main goal was to obtain an accurate knowledge of the spatial and temporal aerosol extent of aerosol concentration and properties for assessing their influence on remote sensed data \cite{bib:holben}. The basic direct solar measurement is based on the Beer-Lambert-Bouger law as:

\begin{equation}
V_{\lambda} =  V^{ext}_{\lambda} d^2 exp(\tau_{\lambda}m) T^{abs}_{\lambda}
\label{eq_beer_lamb_bou}
\end{equation}

where: \\ 
$V_{\lambda}$ = measured voltage. \\ 
$V^{ext}_{\lambda}$ = Extraterrestrial voltage at reference distance $d$=1. \\ 
$\lambda$ = Measured wavelength. \\ 
$d$ = Earth – Sun distance ratio to reference distance. \\ 
$\tau_{\lambda}$ = Total optical depth. \\ $m$ = Optical air mass. \\ 
$T^{abs}_{\lambda}$ = Transmission of absorbing gases.

Further information about the instruments and the measurements can be found in \cite{bib:holben2}.

\section{The Aeronet CASLEO Station}

\begin{figure}[h!]
\centering
\includegraphics[width=0.4\textwidth]{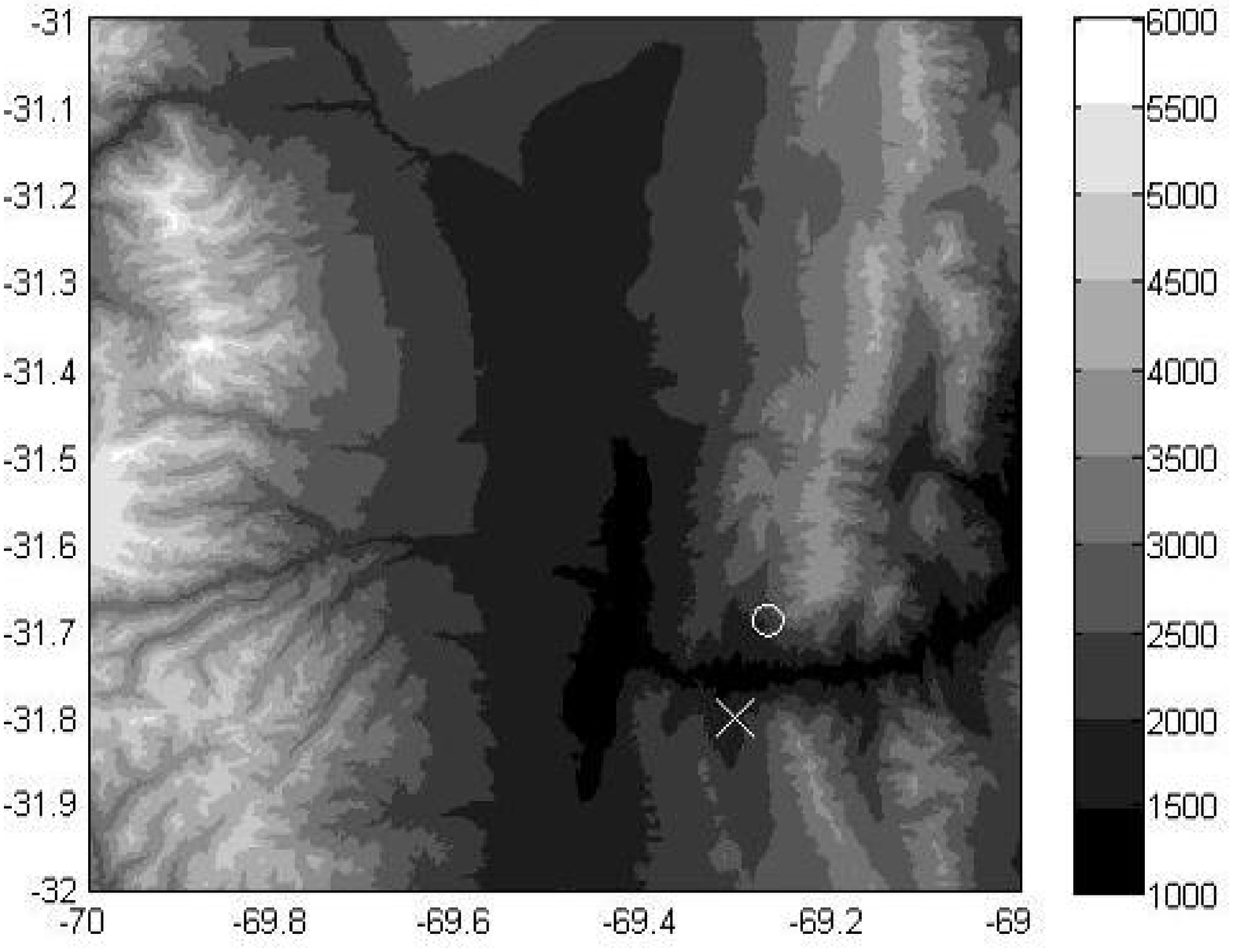}
\caption{Shows with an “X” the Aeronet station, and with an “O” the CTA candidate site location.}
\label{imagen_sat_fig}
\end{figure}

\begin{figure}[h!]
\centering
\includegraphics[width=0.4\textwidth]{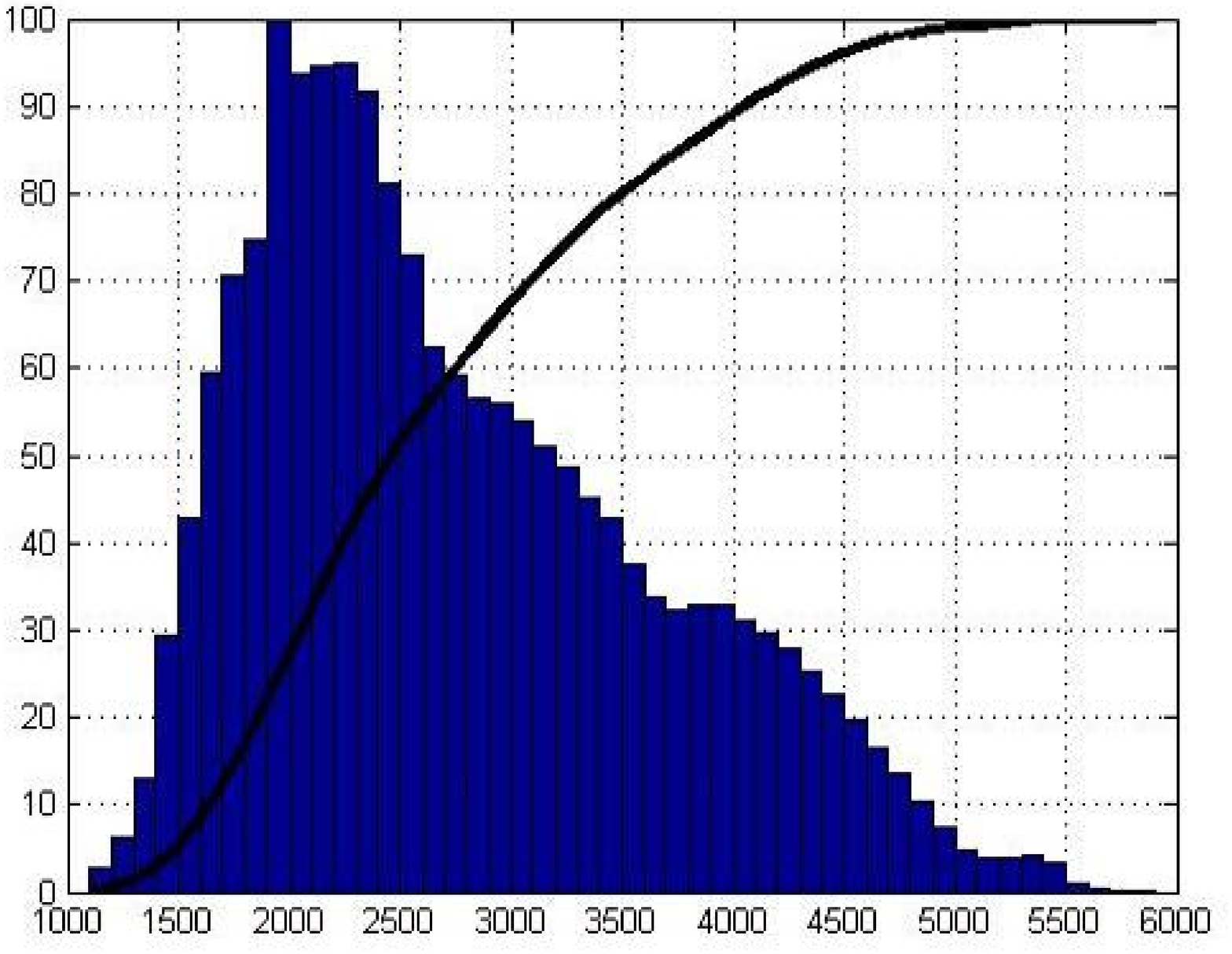}
\caption{Shows the height distribution and the cumulative height distribution of Figure \ref{imagen_sat_fig}.}
\label{histo_fig}
\end{figure}

In a first analysis of the CTA candidate sites, AOD results from satellite borne instruments showed unexpected high values which were neither compatible with the region nor the main goals of an Optical Astronomical Observatory. This phenomenon is explained by the presence of the Andes foothills. This complex topography leads to large temporal and spatial changes in the surface albedo which is critical to retrieve the satellite AOD \cite{bib:cinco}. These changes are due to the presence of ground snow and ice, bright reflecting surfaces and the vegetation canopy seasonal variation.
The complex topography of CASLEO is evidenced on Figures \ref{imagen_sat_fig} \ref{histo_fig} which shows the height distribution of the pixel used to retrieve MODIS Terra and Aqua Level-3 Daily 1 x 1 degree product. This map created from a 52 m x 52 m digital elevation model from \cite{bib:seis} shows that 51.6\% of the observed surface is between 1000 m asl and the Observatory height and 48.4\% is higher up to 5800 m asl.

\begin{figure}[h!]
\centering
\includegraphics[width=0.4\textwidth]{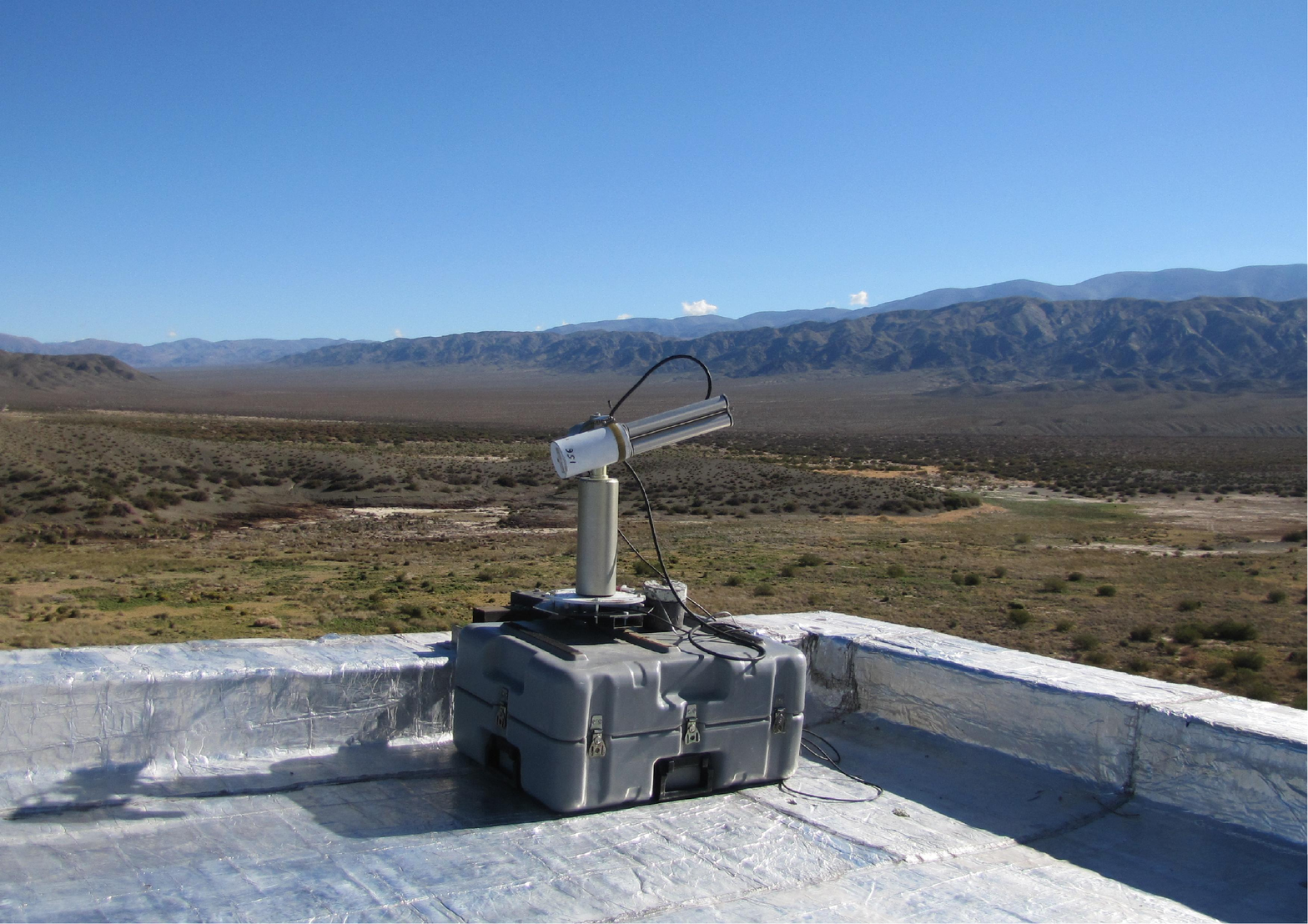}
\caption{Aeronet CASLEO sunphotometer. Installed at the rooftop of the Observatory it has been operating since the beginning of 2011. The instrument monitors the AOD of this location and sends the measured data to the Aeronet database by a satellite uplink to the geosynchronous GOES-E using the antenna behind the instument.}
\label{foto_casleo_fig}
\end{figure}

Aeronet CASLEO station (see Figure \ref{foto_casleo_fig})  is located at the El Leoncito Observatory at 31$^\circ$47’56” S, 69$^\circ$18’24” W at 2552 m asl. This location has been proposed as a South American Aeronet calibration station (as Mauna Loa is for most of the worldwide Aeronet instruments or Izaña-Tenerife is for the European Instrumentation) and chosen by Aeronet NASA and the Lidar Division from CEILAP to provide reliable information about the aerosol load at this CTA candidate site. The Aeronet station has been operational since January 11, 2012 and performing AOD measurements since then.

\section{Measurements}

\begin{figure}[h!]
\centering
\includegraphics[width=0.4\textwidth]{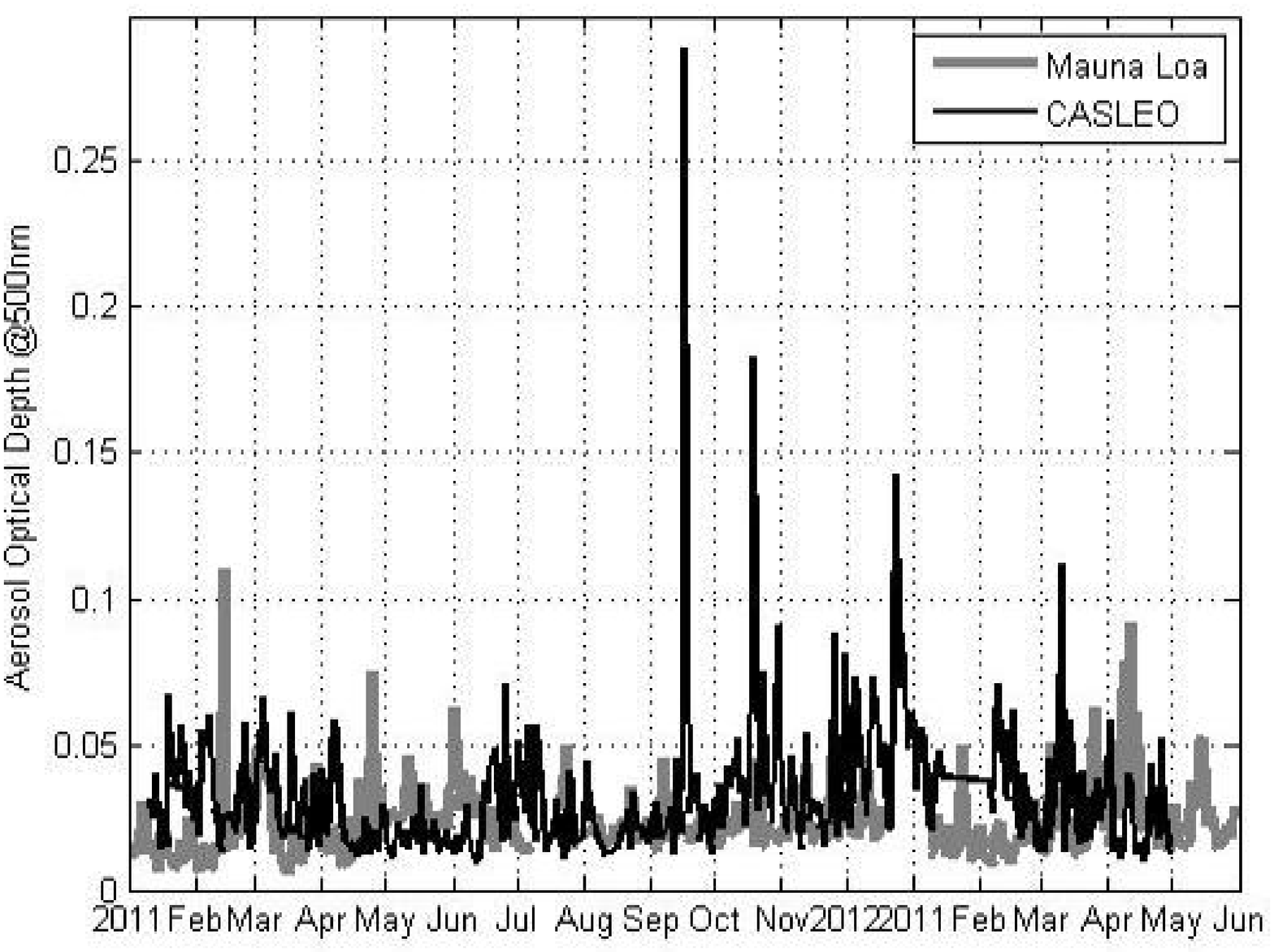}
\caption{Time series functions of the AOD measured at 500 nm for Mauna Loa and CASLEO sites.}
\label{time_series_fig}
\end{figure}

\begin{figure}[h!]
\centering
\includegraphics[width=0.4\textwidth]{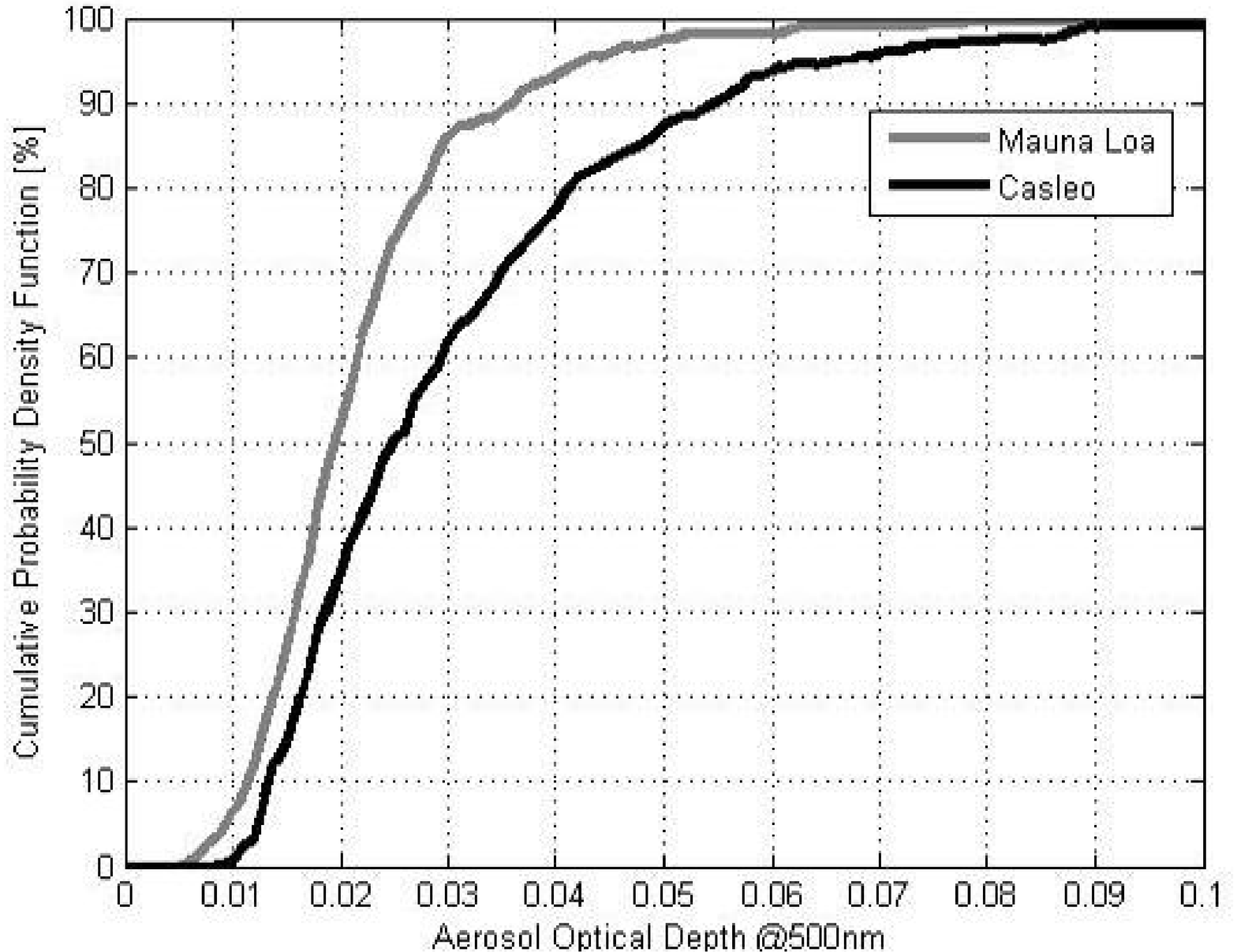}
\caption{Cumulative density functions of the AOD measured at 500 nm for Mauna Loa and CASLEO sites. }
\label{cumulative_fig}
\end{figure}

The Aeronet dataset used to analyze the CASLEO location had 434 days (and 19007 measurements) of quality assured AOD measurements from January 11, 2011 to May 30, 2012. The used wavelengths are 1020 nm, 870 nm, 675 nm, 500 nm, 440 nm, 380 nm and 340 nm. The daily mean AOD values measured at 500 nm of CASLEO are presented on Figures \ref{time_series_fig} \ref{cumulative_fig} compared with Mauna Loa.

As it can be seen from the figures, most of the AOD from both sites are far below 0.05 (92.8\% and 86.9\% of the measured daily means at Mauna Loa and CASLEO respectively). Considering that this condition means a vertical transmission due to the presence of aerosols of 95\% (93\% on 45 deg slant path observation) it can be considered that the radiation loss due to the presence of aerosols is negligible.

\section{Conclusions}

Using the Aeronet CASLEO Level 2 information it can be found that the AOD measured at 500 nm is below 0.05 in 87.9\% of the cases (86.9\% of the daily means), below 0.75 in 96.5\% of the cases (96.8\% of the daily means) and below 0.1 in 98.7\% of the cases (99.2\% of the daily means). This means that only four days during this period have daily mean values above 0.1.
These optical depths correspond to zenithal transmission 95.1\%, 98.8 and 90.5\% from a 100\% meaning an aerosol free atmosphere. For a slant path of 45% the transmission values are 93.2\%, 89.9\% and 86.8\% respectively.
Mauna Loa and CASLEO observation stations are presented showing the potential of the Argentinean site to be a future candidate to perform calibrations for the region.

\vspace*{0.5cm}

\footnotesize{{\bf Acknowledgment:}{The authors would like to thank Prof. \\Brent Holben for establishing and maintaining Mauna Loa site, for his agreement to install an Aeronet unit at CASLEO after our requirement and for his involvement to preserve the data quality and to sustain the operation of Argentinean Aeronet stations. They would also like to thank the CASLEO astronomical complex and specially their technical support team Rodolfo Godoy and Luis Aballay. We gratefully acknowledge support from the following agencies and organizations: Ministerio de Ciencia, Tecnolog\'ia e Innovaci\'on Productiva (MinCyT), Comisi\'on Nacional de Energ\'ia At\'omica (CNEA) and Consejo Nacional  de Investigaciones Cient\'ificas y T\'ecnicas (CONICET) Argentina; State Committee of Science of Armenia; Ministry for Research, CNRS-INSU and CNRS-IN2P3, Irfu-CEA, ANR, France; Max Planck Society, BMBF, DESY, Helmholtz Association, Germany; MIUR, Italy; Netherlands Research School for Astronomy (NOVA), Netherlands Organization for Scientific Research (NWO); Ministry of Science and Higher Education and the National Centre for Research and Development, Poland; MICINN support through the National R+D+I, CDTI funding plans and the CPAN and MultiDark Consolider-Ingenio 2010 programme, Spain; Swedish Research Council, Royal Swedish Academy of Sciences financed, Sweden; Swiss National Science Foundation (SNSF), Switzerland; Leverhulme Trust, Royal Society, Science and Technologies Facilities Council, Durham University, UK; National Science Foundation, Department of Energy, Argonne National Laboratory, University of California, University of Chicago, Iowa State University, Institute for Nuclear and Particle Astrophysics (INPAC-MRPI program), Washington University McDonnell Center for the Space Sciences, USA. The research leading to these results has received funding from the European Union's Seventh Framework Programme ([FP7/2007-2013] [FP7/2007-2011]) under grant agreement nÂ° 262053.}}

\end{document}